# Influence of the pinhole size on the resolution of the 4Pi' microscope studied by means of the optical transfer function

Nicolas Sandeau[†], Hugues Giovannini[*]

*Institut Fresnel UMR 6133 CNRS, Université Paul Cézanne Aix-Marseille III 13397 Marseille cedex 20, France*

**Abstract**

Optical transfer functions are presented for the 4pi' microscope. The effects of the size of the confocal pinhole on the three-dimensional resolution are studied. It is shown that the resolution of the 4Pi' microscope depends weakly on the pinhole size. This result suggest that the pinhole can be removed in order to increase the signal-to-noise ratio.

## 1. Introduction

Strong efforts have been made for increasing the resolution of fluorescence confocal microscopes. Indeed, fluorescence microscopes give precious information about biomedical issues where the volume of observation has to be as small as possible. Among the solutions proposed, the 4Pi microscope [1,2] seems to be the most powerful tool for reducing the axial extent of the investigation volume in fluorescence experiments. For example Egner et al. used this type of microscope to image the Golgi apparatus in live mammalian cells [3] with a ~100 nm axial resolution. Recently, a particular arrangement, referred as 4Pi' microscope [4], has been proposed for increasing the lateral resolution of 4Pi microscopes while maintaining the axial resolution unchanged. The 4Pi' arrangement is obtained from the 4Pi microscope of [1] by adding an image inversion system in one arm of the interferometer. This system allows creating from a point source – typically a fluorophore – a second virtual source. These two sources are symmetric with respect to the common focus of the objective microscopes. The two beams emitted by the sources travel across the two arms of the interferometer and are superimposed on the confocal pinhole. The rules of geometrical optics show that the two images of the sources are symmetric with respect to the centre of the pinhole. The superposition of these two images results in a reduction of the lateral extent of the Molecular Detection Efficiency Function (MDEF).

Interpretations of this effect have been given in terms of geometrical optics [4]. In order to give a more detailed interpretation and for studying the influence of the pinhole size on the resolution, we have computed the optical transfer function (OTF) of the 4Pi', of the 4Pi and of the confocal microscopes. Indeed, it has been shown [5,6] that the OTF allows to understand the influence of microscope parameters on the resolution and to study the properties of new types of microscopes like Teta [7] or 4Pi [8] microscopes.

## 2. Theory

In fluorescence confocal microscopy, the MDEF is determined by the product of the Excitation Efficiency Function (EEF) and of the Collection Efficiency Function (CEF). In a one-photon excitation mode, the EEF describes the intensity of the focused pump beam which illuminates the sample. If we don't consider the polarization of the pump beam, the EEF is obtained with a modified theory [9] of Debye and Pitch [10]. The CEF is the intensity received by the photodetector situated behind the pinhole. The CEF depends on the position of the fluorophore. Numerical calculations of the MDEF have been made with a one-photon excitation [4] and a two-photon excitation [11] regimes by using the approach developed by Richards and Wolf [12] and later adapted by Enderlein [13,14].

The 3D-OTF in fluorescence confocal microscopes is given by the Fourier Transform $\mathcal{F}$ of the MDEF [15]:

$$\begin{aligned}\mathrm{OTF}(k_x, k_y, k_z) &= \mathcal{F}[\mathrm{MDEF}(x,y,z)] \\ &= \mathcal{F}[\mathrm{EEF}(x,y,z)] \otimes \mathcal{F}[\mathrm{CEF}(x,y,z)]\end{aligned} \quad (1)$$

———
[†] Tel.: +33 147 40 55 54; fax: +33 147 40 55 67; e-mail: nicolas.sandeau@ida.ens-cachan.fr
[*] Tel.: +33 491 28 80 66; fax: +33 491 28 80 67; e-mail: hugues.giovannini@fresnel.fr

where z is the coordinate along the optical axis, and x and y are the coordinates in the transverse plane. As the polarisation of the pump beam is not taken into account, the MDEF is symmetric around the optical axis. Thus the OTF is two-dimensional, described by the 2D-OTF with:

$$\text{OTF}(k_r, k_z) = \mathcal{F}[\text{MDEF}(r,z)] = \iint \text{MDEF}(r,z) \cdot e^{-i(k_r \cdot r + k_z \cdot z)} dr\, dz \quad (2)$$

where r is represents the position of the fluorophore in the transverse plane. Thus, in one dimension we have:

$$\text{OTF}(k_r) = \mathcal{F}[\text{MDEF}(r)]$$
$$\text{OTF}(k_z) = \mathcal{F}[\text{MDEF}(z)] \quad (3)$$

where $\text{OTF}(k_r)$ and $\text{OTF}(k_z)$ are respectively the transverse and the longitudinal optical transfer functions.

## 3. Numerical simulations

### 3.1. Lateral resolution

We have calculated the MDEF and the OTF of the 4Pi' microscope for different pinhole diameters φ. In the calculations, the excitation wavelength $\lambda_{exc}$=488 nm, the emission wavelength $\lambda_{fluo}$=525 nm. These conditions correspond to the practical case of an argon-ion used as the pump source and Oregon Green (Molecular Probes, Eugene, OR) fluorophores. The objectives are immersion-oil (n=1.518) objectives with a numerical aperture NA=1.3. The magnification m=40 and the incident pump beam is a Gaussian beam with β=1 [16]. These conditions correspond to practical cases and are typical of confocal microscopes used in fluorescence experiments. With these characteristics, the diameter of the Airy function on the photodetector is equal to 19.7 μm. Figure 1 shows the section of the MDEF in the common focal plane of the two microscope objectives calculated for different pinhole diameters φ: one well adapted to the Airy spot (φ=20 μm), another smaller (φ=10 μm), and another much larger (φ=40 μm). A comparison is drawn between the MDEF of the 4Pi microscope and the MDEF of the 4Pi' microscope. As expected, the maximum value of the MDEF increases when φ increases. Reducing the pinhole size leads to a weaker signal-to-noise ratio. In all the cases, the lateral extent of the MDEF is smaller for the 4Pi' microscope than for the 4Pi microscope. This result shows that a better lateral resolution is obtained with the 4Pi' microscope. When φ tends to 0, the two systems become equivalent. In order to discuss more precisely on the origin of the improvement of resolution obtained with the 4Pi' microscope, we have computed the OTF. The results are represented in Figure 2. For comparing the shapes of the OTF obtained for different values of φ, we have normalized the results. One can see that, for the same value of φ, the OTF of the 4Pi' microscope is broader than the OTF of the 4Pi microscope. In particular, the amplitude of the OTF of the 4Pi' microscope is stronger at high frequencies. This is the main cause of the improvement of the lateral resolution obtained with the 4Pi' microscope. Moreover one can see that the shape of the OTF of the 4Pi' microscope depends weakly on the pinhole size. This result suggests that the pinhole size has little influence on the lateral resolution of the 4Pi' microscope. This result is confirmed (see Table 1) by the calculations of the lateral width of the MDEF.

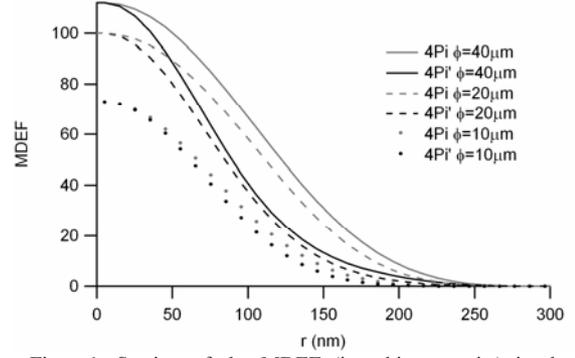

Fig. 1. Section of the MDEF (in arbitrary units) in the common focal plane of the objectives of the 4Pi and the 4Pi' microscope.

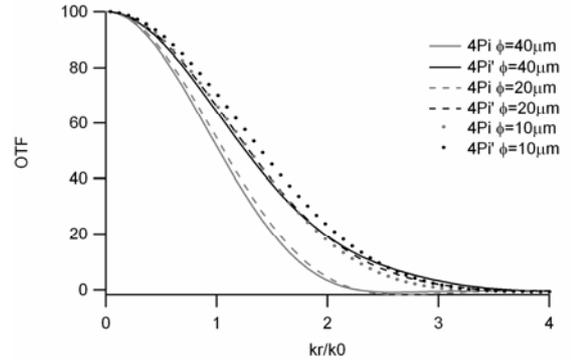

Fig. 2. Section of the normalized transverse OTF (in arbitrary units) in the common focal plane of the objectives of a 4Pi and a 4Pi' microscope.

Table 1. Width of the normalized lateral MDEF of 4Pi and 4Pi' microscopes at 20% of the maximum as a function of the pinhole diameter φ.

| φ (μm) | 10 | 20 | 40 |
|---|---|---|---|
| 4Pi-Confocal | 129 nm | 159 nm | 165 nm |
| 4Pi' | 119 nm | 127 nm | 127 nm |

## 3.2. Axial resolution

In Figure 3 we have represented the MDEF along the optical axis. Calculations have been made for the confocal microscope and for the 4Pi' microscope. As we neglect the effects of polarization, the MDEF of the 4Pi' microscope is equivalent to that of the 4Pi microscope along the optical axis. One can see the side-lobes obtained with the 4Pi' microscope due to the interferences between the incident pump beams and the interferences between the emitted beams. For the conventional confocal microscope, the axial extent of the MDEF increases when the pinhole size increases. On the other hand, the pinhole size has little influence on the heights of the first side-lobes (see Table 2) obtained with the 4Pi' microscope.

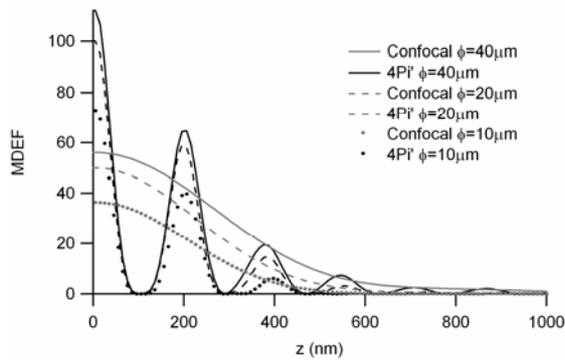

Fig. 3. Section of the MDEF (in arbitrary units) along the optical axis of the objectives of a 4Pi' and a classical confocal microscope.

Figure 4 represents the corresponding axial OTF. As expected, the OTF of the confocal microscope broadens when the pinhole size decreases. The OTF of the 4Pi' microscope exhibits peaks at high frequencies. The origin of the resolution improvement along the optical axis obtained with the 4Pi' microscope, and represented by the width of the central peak of the MDEF, lies on this property. However, although the central peak of the MDEF of the 4Pi' microscope is narrower than that of the confocal microscope, the particular shape of the OTF of the 4Pi' microscope leads to the presence of side-lobes in the MDEF.

Table 2. Amplitude of the side-lobes of the normalized axial MDEF of 4Pi' microscopes as a function of the pinhole diameter φ.

| φ (μm) | 10 | 20 | 40 |
| --- | --- | --- | --- |
| 1st side-lobes | 55 % | 58 % | 58 % |
| 2nd side-lobes | 9 % | 14 % | 18 % |

## 4. Conclusion

All these results suggest that the pinhole size has little influence on the resolution of the 4Pi' microscope contrary to classical 4Pi and confocal microscopes. As increasing the pinhole size leads to an improvement of the signal-to-noise ratio, this property is a major advantage of the 4Pi' microscope over the 4Pi microscope and the classical confocal microscope. Thus, 4Pi' microscopes can be used with a large pinhole or even without any pinhole to observe weakly fluorescent biological samples.

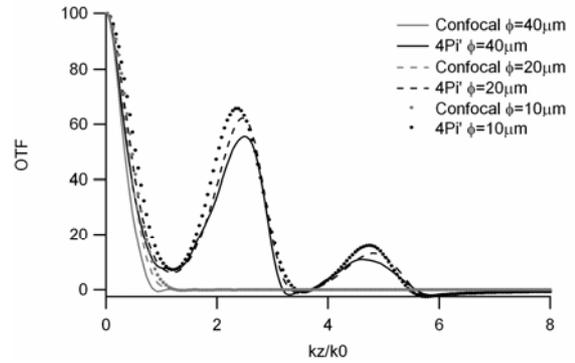

Fig. 4. Section of the normalized axial OTF (in arbitrary units) along the optical axis of the objectives of a 4Pi' and a classical confocal microscope.